\documentclass[showpacs,amssymb,aps]{revtex4}

\newcommand{\be}{\begin{equation}}
\newcommand{\ee}{\end{equation}}

\newcommand{\un}{${\rm U}(N)\;$}

\usepackage{graphicx}

\begin{document}

\title{Dispersion relations for the self-energy in non-commutative field theories} 

\author{F. T. Brandt$^a$, Ashok Das$^b$, J. Frenkel$^a$}
\affiliation{$^a$ Instituto de F\'{\i}sica,
Universidade de S\~ao Paulo,
S\~ao Paulo, SP 05315-970, BRAZIL}
\affiliation{$^b$Department of Physics and Astronomy,
University of Rochester,
Rochester, NY 14627-0171, USA}

\bigskip

\begin{abstract}

We study the IR/UV connection in non-commutative $\phi^{3}$ theory as
well as in non-commutative QED from the point of view of the dispersion
relation for the self-energy. We show that, although the imaginary
part of the self-energy is well behaved as the parameter of
non-commutativity vanishes, the real part becomes divergent as a
consequence of the high energy behavior of the dispersion
integral. Some other interesting features that arise from this analysis
are also briefly discussed.

\end{abstract}


\maketitle

\section{Introduction}
In recent years, non-commutative field theories have been studied from
various points of view
\cite{Gonzalez-Arroyo:1983ub,Filk:1996dm,Martin:1999aq,Sheikh-Jabbari:1999iw,Krajewski:1999ja,Bigatti:1999iz,Maldacena:1999mh,Iso:2000ew,Arcioni:1999hw,Minwalla:1999px,Aref'eva:1999sn,Gross:2000ba,Matusis:2000jf,Hayakawa:1999yt,girotti:2000gc,Zanon:2000nq,Das:2001kf,Khoze:2000sy,Liu:2000ad,VanRaamsdonk:2001jd,Bonora:2000ga,Armoni:2000xr,Ruiz:2000hu,Bichl:2001cq,Szabo:2001kg,Douglas:2001ba}. 
These are theories defined on a manifold where the coordinates do not
commute, rather satisfy 
\begin{equation}
[x^{\mu},x^{\nu}] = i \theta^{\mu\nu}
\end{equation}
Here, $\theta^{\mu\nu}$ is a constant anti-symmetric tensor and, for
unitarity to hold, one normally assumes that $\theta^{0i}=0$, namely,
one assumes only the space coordinates to have 
non-commutativity \cite{Gomis:2000zz,Bahns:2002vm}.

In an earlier paper \cite{Brandt:2001ud}, 
we studied the self-energy in non-commutative QED
(as well as in non-commutative  
$\phi^{3}$ theory) in $n$ dimensions and it was shown there
that while the imaginary parts of the self-energy were well behaved as
$\theta^{ij}\rightarrow 0$, the real parts were divergent for $n\geq
4$. This is, in fact, a puzzling behavior. Namely, if causality were
to hold in non-commutative theories (QED as well as the 
$\phi^{3}$ theory), we would expect the real and the imaginary parts
of the self-energy to be related through a dispersion relation and
then, it is not clear how a well behaved imaginary part can lead to a
divergent structure in the real part. It is with this in mind that we
have undertaken a systematic study of the dispersion relation in the
non-commutative $\phi^{3}$ theory as well as QED in $n$-dimensions and our
study leads to various interesting results which we discuss in this paper.

We note that a dispersion relation, which is a statement about
causality in a quantum field theory, holds for any well behaved
analytic function that vanishes at infinity \cite{Eden:1966,Sakurai:1967}.
This is particularly true
for the retarded self-energy which is expected to satisfy
\begin{equation}
{\rm Re}\,\Pi^{\rm (R)} (p^{0},\vec{p}) = \frac{1}{\pi}\,
{\rm P}\,\int_{-\infty}^{\infty} d\omega\,\frac{{\rm Im}\,\Pi^{\rm (R)}
(\omega,\vec{p})}{\omega - p^{0}}
\end{equation}
where \lq\lq P'' stands for the Cauchy principal value. Furthermore,
using the  fact that the imaginary part of the retarded
self-energy is odd in the energy variable, the dispersion relation can
be equivalently written as
\begin{equation}
{\rm Re}\,\Pi^{\rm (R)} (p^{0},\vec{p}) = \frac{2}{\pi}\,{\rm P}\,
\int_{0}^{\infty} d\omega\,\frac{\omega\, {\rm Im}\,\Pi^{\rm (R)}
(\omega,\vec{p})}{\omega^{2}-(p^{0})^{2}}\label{dispersion}
\end{equation}
This form of the dispersion relation involves only positive values of
the energy inside the integral and, therefore, is quite useful from our
point of view. The reason for this is quite clear. In general, the
Feynman amplitudes are not expected to satisfy a simple dispersion relation,
but as we will show in section {\bf II}, for positive energy, the
Feynman and the retarded self-energies coincide and, consequently,
satisfy (\ref{dispersion}). In the subsequent
discussions, we will omit the principal value symbol for simplicity,
although it is to be understood throughout.

In a non-commutative theory, the self-energy has a planar term and a
non-planar term. The behavior of the planar terms is the same as in a
commutative theory and, therefore, we concentrate only on the
non-planar terms in the self-energy, commenting where necessary on the
planar terms. It is in the non-planar terms that the nontrivial phase
factors of the interaction vertices are likely to modify the complex
structure of the amplitudes and, therefore, the validity of the
dispersion relations is more crucial for these terms. We note that the
non-planar terms are ultraviolet finite (unlike the planar ones) and,
therefore, may satisfy an unsubtracted dispersion relation. Furthermore, 
it is also in these terms that the real parts of the amplitudes show
a divergence as $\theta^{ij}\rightarrow 0$ which we would like to
understand from the point of view of the dispersion relations. 
We assume, as is conventionally done, that $\theta^{0i}=0$ because it is
in this case that explicit calculations exist and the theories are
believed  to be unitary. In
section {\bf II}, we evaluate the imaginary part of the Feynman
self-energy in the massless non-commutative $\phi^{3}$ theory. We
show that, for positive values of the energy, the Feynman amplitude
coincides  with
the retarded one and use the dispersion relation to calculate the real
part of the self-energy. We, then, show that this coincides exactly
with the real part of the self-energy calculated 
in \cite{Brandt:2001ud} proving that
the dispersion relation indeed holds true in the non-commutative theory.
It follows from this
analysis that the divergence structure of the real part of the
self-energy, as $\theta^{ij}\rightarrow 0$, arises from large
values of energy, $\omega$, in the dispersion integral, giving yet
another  manifestation of the IR/UV
mixing. In section {\bf III}, we carry out a similar analysis for QED
and show that the dispersion relations hold in this case as well. In
section {\bf IV}, we show that the real and the imaginary parts of the
self-energy, both for the non-commutative
$\phi^{3}$ theory as well as for non-commutative QED, can be
expressed in closed form (without any Feynman parameter integrals) and
this brings out some nice features. In section {\bf V}, we discuss
various other interesting features that arise from our analysis. Some
mathematical details are discussed briefly in the appendix. 

\section{Dispersion relation in the non-commutative $\phi^{3}$  theory}

In this section, we will discuss the dispersion relation for the
self-energy in the massless non-commutative $\phi^{3}$ theory in
$n$-dimensions. Since the dispersion relation for the planar part of the
self-energy is well understood, we will concentrate here only on the
non-planar part of the amplitude. We note that the non-planar
part of the self-energy, in $n$-dimensions, has the form (apart from a
factor of $\frac{\lambda^{2}}{4}$ where $\lambda$ represents the
scalar coupling)
\begin{equation}
i\Pi^{\rm (non\hbox{-}planar)} (p)  = \int
\frac{d^{n}k}{(2\pi)^{n}}\,\frac{e^{i\bar{\theta}\cdot
k}}{(k^{2}+i\epsilon)((k+p)^{2} + i\epsilon)}
\end{equation}
Here, we have defined
\begin{equation}
\bar{\theta}^{\mu} = \theta^{\mu\nu} p_{\nu}
\end{equation}
which has only a nontrivial space component. The integral over $k^{0}$
can be trivially done yielding
\begin{equation}
\Pi^{\rm (non\hbox{-}planar)} (p) = \frac{1}{4 (2\pi)^{n-1}} \int d^{n-1}k\,
\frac{\cos \bar{\theta}\cdot k}{|\vec{k}||\vec{k} + \vec{p}|}
\left[\frac{1}{p^{0} + |\vec{k}| + |\vec{k}+\vec{p}| - i\epsilon} -
\frac{1}{p^{0} - |\vec{k}| - |\vec{k}+\vec{p}| +
i\epsilon}\right]\label{feynman} 
\end{equation}
It follows now that the imaginary part of the non-planar self-energy
has the form
\begin{equation}
{\rm Im}\, \Pi^{\rm (non\hbox{-}planar)} (p) = \frac{\pi}{4 (2\pi)^{n-1}}
\int d^{n-1} k\,\frac{\cos \bar{\theta}\cdot
k}{|\vec{k}||\vec{k}+\vec{p}|} \left(\delta (p^{0}-
|\vec{k}|-|\vec{k}+\vec{p}|) + \delta (p^{0} + |\vec{k}| +
|\vec{k}+\vec{p}|)\right)\label{imaginary}
\end{equation}

There are several things to note here. First, in spite of the
similarities of the non-commutative theories to thermal field theories
as alluded to in \cite{Brandt:2001ud}, we see that the imaginary part
of the self-energy
only involves two delta functions as opposed to four, normally
encountered in a thermal field 
theory \cite{kapusta:book89,lebellac:book96,das:book97}. 
This is another evidence of the
fact that, in spite of the non-analyticity present in non-commutative
theories, there are no additional channels of reaction (as would be the case
in thermal field theories) and, therefore, no new branch cuts. Second, for positive
energy, $p^{0} > 0$ (which is what is needed in the dispersion
relation (\ref{dispersion})), only one of the delta functions contributes to
the imaginary part. Without giving the mathematical details (which we
discuss in the appendix), we note that, in this case, the
imaginary part of the self-energy can be written as (for $p^{0} > 0$) 
\begin{equation}
{\rm Im}\,\Pi^{\rm (non\hbox{-}planar)} (p) = \frac{\theta (p^{2})
\pi^{\frac{n}{2}+1}}{(2\pi)^{n}} \int_{0}^{1} dx\,
\frac{1}{|M|^{4-n}}\,\left(\frac{|\bar{\theta}||M|}{2}\right)^{2-\frac{n}{2}}\,
J_{\frac{n}{2}-2} (|\bar{\theta}||M|)\label{imaginarypi}
\end{equation}
where we have defined
\begin{equation}
|M| = |(-x(1-x)p^{2})|^{\frac{1}{2}},\qquad |\bar{\theta}| =
(-\bar{\theta}\cdot \bar{\theta})^{\frac{1}{2}}\label{M}
\end{equation}
We note here that the imaginary part of the self-energy, for $p^{0} <
0$, is also exactly the same so that the imaginary part of the Feynman
amplitude in (\ref{imaginary}) is symmetric in $p^{0}$. Furthermore,  
from the identity satisfied by the Bessel functions \cite{gradshteyn},
\begin{equation}\label{bessel}
\lim_{z\rightarrow 0}\,z^{-\nu} J_{\nu} (z) \rightarrow \frac{1}
{2^{\nu}\,\Gamma(\nu+1)}
\end{equation}
it follows that the imaginary part of the self-energy is well behaved as
$|\bar{\theta}|\rightarrow 0$.

Let us recall that the self-energy for the non-commutative
$\phi^{3}$ theory, in $n$ dimensions, was evaluated earlier in \cite{Brandt:2001ud} and
the non-planar part was shown to have  the form 
\begin{equation}
\Pi^{\rm (non\hbox{-}planar)} (p) = \frac{2\pi^{\frac{n}{2}}}{(2\pi)^{n}}
\int_{0}^{1} dx\, \frac{1}{(M^{2})^{2-\frac{n}{2}}}\,
\left(\frac{|\bar{\theta}|M}{2}\right)^{2-\frac{n}{2}}\,
K_{2-\frac{n}{2}} (|\bar{\theta}|M)\label{amplitude}
\end{equation}
Using the series representations for the Bessel functions, it can be
checked that the imaginary part of this expression indeed coincides
exactly with expression (\ref{imaginarypi})
following from the direct evaluation of the imaginary part. 

So far, we have been working with the time-ordered Feynman amplitude. The
dispersion relation, on the other hand, holds for the retarded
amplitudes. In fact, we note from the dispersion relation in
(\ref{dispersion}) that
we need the imaginary part of the retarded self-energy only for
positive values of energy. We show now that, for positive energy, the retarded
self-energy coincides with the time-ordered Feynman self-energy that we have
already evaluated. This is easily seen as follows. From the definition
of the retarded and the time-ordered amplitudes, it follows that
\begin{equation}
{\rm R}\,(\phi(x_{1})\phi(x_{2})) = {\rm T}\,(\phi(x_{1})\phi(x_{2}))
- \phi(x_{2})\phi(x_{1})\label{relation}
\end{equation}
Furthermore, from the definition of the positive energy Green's
function, we know that we can write
\begin{equation}
\langle 0|\phi(x_{1})\phi(x_{2})|0\rangle = \int d^{n}p\, \theta
(p^{0}) \delta (p^{2})\,e^{-ip\cdot (x_{1}-x_{2})}\,f(p)
\end{equation}
from which it follows that
\begin{equation}
\langle 0|\phi(x_{2})\phi(x_{1})|0\rangle = \int
d^{n}p\,\theta(-p^{0}) \delta (p^{2})\, e^{-ip\cdot
(x_{1}-x_{2})}\,f(-p)
\end{equation}
Therefore, we see that the vacuum expectation value of
the second term on the right hand side in
(\ref{relation})  has
contributions only for negative values of the energy. 
As a result, for positive
energy, this term does not contribute and the retarded and the Feynman
self-energies coincide. As a practical rule for obtaining the retarded
self-energy from the Feynman amplitude in (\ref{feynman}), we note
that the  Feynman $i\epsilon$'s in (\ref{feynman}) can be replaced
with  $p^{0}\rightarrow
p^{0}+i\epsilon$ to yield the retarded self-energy. In
this case, it follows from (\ref{feynman}) that the imaginary part of
the  retarded
self-energy is anti-symmetric in $p^{0}$, as it should be.
Furthermore, the real parts of the time-ordered and the retarded
self-energy are the same for all energies. 

With this, we can now use the dispersion relation,
Eq. (\ref{dispersion}), to determine the real part of the self-energy. 
\begin{eqnarray}
{\rm Re}\,\Pi^{\rm (non\hbox{-}planar)} (p) & = & \frac{2}{\pi}
\int_{0}^{\infty} d\omega\,\frac{\omega\, {\rm Im}\,\Pi^{\rm
(non\hbox{-}planar)} (\omega,\vec{p})}{\omega^{2}-(p^{0})^{2}}\nonumber\\
 & = & \frac{2\pi^{\frac{n}{2}}}{(2\pi)^{n}} \int_{0}^{1} dx
\int_{0}^{\infty} 
d\omega\,\frac{\omega\,\theta(\omega^{2}-\vec{p}^{2})}{\omega^{2} -
(p^{0})^{2}}\,\frac{1}{|M(\omega,\vec{p})|^{4-n}}\,\left(\frac{|\bar{\theta}|
|M(\omega,\vec{p})|}{2}\right)^{2-\frac{n}{2}}
J_{\frac{n}{2}-2} (|\bar{\theta}||M(\omega,\vec{p})|)\label{realpi}
\end{eqnarray}
It is worth noting here that, as $|\bar{\theta}|\rightarrow 0$, even
though the combination of the Bessel functions is well behaved (see
(\ref{bessel})),  the integral diverges as (for $n>4$)
\begin{equation}
{\rm Re}\,\Pi^{\rm (non\hbox{-}planar)} \rightarrow \omega^{n-4} \sim
|\bar{\theta}|^{4-n}
\end{equation}
where we have used dimensional reasoning in the last
relation. Thus, even though the imaginary part of the self-energy is
well behaved as $|\bar{\theta}|\rightarrow 0$, the dispersion relation
induces a divergence in the real part of the amplitude which arises
from the region of integration involving large values of energy. This
is yet another interesting manifestation of the IR/UV mixing. We will
comment more on this later in the discussions.

For the present, let us simply note that with a simple change of
variables, we can write the real part of the self-energy to be
\begin{eqnarray}
{\rm Re}\,\Pi^{\rm (non\hbox{-}planar)} (p) & = &
\frac{2\pi^{\frac{n}{2}}}{(2\pi)^{n}}
\left(\frac{|\bar{\theta}|}{2}\right)^{2-\frac{n}{2}} \int_{0}^{1}
dx\,(x(1-x))^{\frac{n}{4}-1} \int_{0}^{\infty} dz\,\frac{1}{z^{2} -
p^{2}}\,z^{\frac{n}{2}-1} J_{\frac{n}{2}-2}
(|\bar{\theta}|\sqrt{x(1-x)}z)\nonumber\\
 & = & \frac{2\pi^{\frac{n}{2}}}{(2\pi)^{n}}\, {\rm Re}\, \int_{0}^{1}
dx\,\frac{1}{(M^{2})^{2-\frac{n}{2}}}\,\left(\frac{|\bar{\theta}|M}{2}
\right)^{2-\frac{n}{2}} K_{2-\frac{n}{2}} (|\bar{\theta}|M)\label{realpi'}
\end{eqnarray}
where 
\begin{equation}
M^{2} = -x(1-x)p^{2}
\end{equation}
and we have used standard integrals involving Bessel 
functions \cite{gradshteyn}.
We can now compare this real part of the amplitude determined from the
dispersion relation with that from the exact amplitude in
(\ref{amplitude})  calculated
earlier and we see that they agree completely.

\section{Dispersion relation in non-commutative QED}

We can now study the dispersion relation in non-commutative
QED. As we have already shown in \cite{Brandt:2001ud}, the
self-energy  for the photon is transverse and, in a general covariant
gauge,  can be parameterized as
\begin{equation}
\Pi_{\mu\nu} = (\eta_{\mu\nu} p^{2} - p_{\mu}p_{\nu})\, \Pi_{1} +
\frac{\bar{\theta}_{\mu}\bar{\theta}_{\nu}}{\bar{\theta}^{2}}\,\Pi_{2}
\label{decomposition}
\end{equation}
We note that $\Pi_{1},\Pi_{2}$ in (\ref{decomposition}) are defined
such that  $\Pi_{1} = \frac{A}{p^{2}},\Pi_{2}=B$ of ref. \cite{Brandt:2001ud}. Once again,
we will only concentrate on the non-planar parts of the self-energy
and will comment on the planar part later.

Let us first analyze the dispersion relation for $\Pi_{1}^{\rm
(non\hbox{-}planar)}$ before discussing the same for $\Pi_{2}^{\rm
(non\hbox{-}planar)}$. The imaginary part of $\Pi_{1}^{\rm
(non\hbox{-}planar)}$ can 
be directly calculated from the Feynman diagram, as in the last
section, and, in this case, leads to 
\begin{eqnarray}
{\rm Im}\, \Pi_{1}^{\rm (non\hbox{-}planar)} & = &
\frac{(-1)^{\frac{n}{2}}e^{2}\pi^{\frac{n}{2}+1}\theta(p^{2})}{(2\pi)^{n}}
\int_{0}^{1} dx\,\frac{1}{|M|^{4-n}}\left[a_{1}
\left(\frac{|\bar{\theta}||M|}{2}\right)^{2-\frac{n}{2}}
J_{\frac{n}{2}-2} (|\bar{\theta}||M|)\right. \nonumber\\
 &  & \left. \qquad\qquad \;\;\;\;\;\;\;\;\;\;\;\;\;\;\;\;\;\;\;\;\;
\;\;\;\;\;\;\;\;\;\;\;\;\;\;\;
+ 2\,a_{2}\left(\frac{|\bar{\theta}||M|}{2}\right)^{3-\frac{n}{2}}
J_{\frac{n}{2}-3} (|\bar{\theta}||M|)\right]
\end{eqnarray}
Here, we have defined
\begin{eqnarray}
a_{1} & = & 3 + 2(n-1)x - 4(n-2) x^{2} - 2(1-\xi) x(6x-5)\nonumber\\
a_{2} & = & (1-\xi)\left((1+4x-4x^{2}) - \frac{1-\xi}{4}\right)
\end{eqnarray}
with $\xi$ representing the gauge fixing parameter. As can be easily
checked using (\ref{bessel}),  this imaginary part is well behaved as
$|\bar{\theta}|\rightarrow 0$.

For positive energy, which is involved in the dispersion relation
(\ref{dispersion}), 
as we have already shown, the Feynman and the retarded
amplitudes coincide. As a result, we can evaluate the real part of
$\Pi_{1}^{\rm (non\hbox{-}planar)}$ as
\begin{eqnarray}
{\rm Re}\,\Pi_{1}^{\rm (non\hbox{-}planar)} & = &
\frac{2(-\pi)^{\frac{n}{2}}e^{2}}{(2\pi)^{n}} \int_{0}^{1} dx
\int_{0}^{\infty}
d\omega\,\frac{\omega\,\theta(\omega^{2}-\vec{p}^{2})}{\omega^{2} -
(p^{0})^{2}}
\frac{1}{|M(\omega,\vec{p})|^{4-n}}\nonumber\\
 &  & \times \left[a_{1}\left(\frac{|\bar{\theta}||M|}
{2}\right)^{2-\frac{n}{2}} J_{\frac{n}{2}-2} (|\bar{\theta}||M|) +
2a_{2} \left(\frac{|\bar{\theta}||M|}{2}\right)^{3-\frac{n}{2}}
J_{\frac{n}{2}-3} (|\bar{\theta}||M|)\right]\label{realpione}
\end{eqnarray}
Once again, we see that as $|\bar{\theta}|\rightarrow 0$, the
combination of the Bessel functions is well behaved, but the integral
develops a divergence coming from large values of $\omega$ as (for
$n>4$)
\begin{equation}
{\rm Re}\,\Pi_{1}^{\rm (non\hbox{-}planar)}\rightarrow \omega^{n-4} \sim
|\bar{\theta}|^{4-n} 
\end{equation}
which shows the manifestation of the IR/UV mixing.

We note that there are more structures in (\ref{realpione}) compared
with (\ref{realpi}). Nevertheless, with a simple change of variables,
we can  evaluate
this using standard integrals of Bessel functions to obtain \cite{gradshteyn}
\begin{eqnarray}
{\rm Re}\,\Pi_{1}^{\rm (non\hbox{-}planar)} & = &
\frac{2(-\pi)^{\frac{n}{2}}e^{2}}{(2\pi)^{n}} \int_{0}^{1} dx
\left[a_{1} \left(\frac{|\bar{\theta}|}{2}\right)^{2-\frac{n}{2}}
(x(1-x))^{\frac{n}{4}-1} 
\int_{0}^{\infty} \frac{dz}{z^{2}-p^{2}} z^{\frac{n}{2}-1}
J_{\frac{n}{2}-2} (|\bar{\theta}|\sqrt{x(1-x)}z)\right.\nonumber\\
 &  & \qquad\left. +
2a_{2}\left(\frac{|\bar{\theta}|}{2}\right)^{3-\frac{n}{2}}
(x(1-x))^{\frac{n}{4}-\frac{1}{2}} \int_{0}^{\infty}
\frac{dz}{z^{2}-p^{2}} z^{\frac{n}{2}} J_{\frac{n}{2}-3}
(|\bar{\theta}|\sqrt{x(1-x)}z)\right]\nonumber\\
 & = & {\rm Re}\,\frac{(-\pi)^{\frac{n}{2}}e^{2}}{(2\pi)^{n}}
\int_{0}^{1} dx\,\frac{1}{(M^{2})^{2-\frac{n}{2}}}\left[2a_{1}
\left(\frac{|\bar{\theta}|M}{2}\right)^{2-\frac{n}{2}}
K_{\frac{n}{2}-2} (|\bar{\theta}|M) - 4a_{2}
\left(\frac{|\bar{\theta}|M}{2}\right)^{3-\frac{n}{2}}
K_{\frac{n}{2}-3} (|\bar{\theta}|M)\right]\nonumber\\
 &  & 
\end{eqnarray}
This can, in fact, be checked to coincide completely with the
corresponding term in the  real
part of the exact self-energy calculated in \cite{Brandt:2001ud}.

The discussion of the dispersion relation for $\Pi_{2}$ proceeds in a
similar manner. Therefore, without giving details, we simply note that
the imaginary part of $\Pi_{2}^{\rm (non\hbox{-}planar)}$ has the form (we
note that, as pointed out in \cite{Brandt:2001ud}, $\Pi_{2}$ does not have a planar
part)
\begin{eqnarray}
{\rm Im}\,\Pi_{2}^{\rm (non\hbox{-}planar)} & = &
\frac{(-1)^{\frac{n}{2}}e^{2}\pi^{\frac{n}{2}+1}\theta(p^{2})
p^{2}}{(2\pi)^{n}} \int_{0}^{1} dx\,\frac{1}{|M|^{4-n}}\nonumber\\
 &  & \times \left[-\frac{b_{1}}{2}
\left(\frac{|\bar{\theta}||M|}{2}\right)^{1-\frac{n}{2}}
J_{\frac{n}{2}-1} (|\bar{\theta}||M|) + b_{2}
\left(\frac{|\bar{\theta}||M|}{2}\right)^{2-\frac{n}{2}}
J_{\frac{n}{2}-2} (|\bar{\theta}||M|)\right.\nonumber\\
 &  & \left. 
\;\;
-2\,b_{3}\left(\frac{|\bar{\theta}||M|}{2}\right)^{3-\frac{n}{2}}
J_{\frac{n}{2}-3} (|\bar{\theta}||M|) +
4b_{4}\left(\frac{|\bar{\theta}||M|}{2}\right)^{4-\frac{n}{2}}
J_{\frac{n}{2}-4} (|\bar{\theta}||M|)\right]
\end{eqnarray}
where we have defined
\begin{eqnarray}
b_{1} & = & - 4(n-2)^{2} x(1-x)\nonumber\\
b_{2} & = & 2(3-2n) x(1-x) + 1- 2x^{2} + 2(1-\xi) (n-4) x\nonumber\\
b_{3} & = & (1-\xi) \left(2x - (n-6)
\frac{(1-\xi)}{4}\right)\nonumber\\
b_{4} & = & - \frac{(1-\xi)^{2}}{4}
\end{eqnarray} 

The dispersion relation now leads to
\begin{eqnarray}
{\rm Re}\,\Pi_{2}^{\rm (non\hbox{-}planar)} & = & \frac{2}{\pi}
\int_{0}^{\infty} d\omega\,\frac{\omega\,{\rm Im}\,\Pi_{2}^{\rm
(non\hbox{-}planar)}(\omega,\vec{p})}{\omega^{2} -
(p^{0})^{2}}\nonumber\\
 & = & {\rm Re}\,\frac{e^{2}(-\pi)^{\frac{n}{2}}p^{2}}{(2\pi)^{n}}
\int_{0}^{1} dx\,\frac{1}{(M^{2})^{2-\frac{n}{2}}}\left[b_{1}
\left(\frac{|\bar{\theta}|M}{2}\right)^{1-\frac{n}{2}}
K_{\frac{n}{2}-1} (|\bar{\theta}|M) + 2b_{2}
\left(\frac{|\bar{\theta}|M}{2}\right)^{2-\frac{n}{2}}
K_{\frac{n}{2}-2} (|\bar{\theta}|M)\right.\nonumber\\
 &  & \left. + 4b_{3}
\left(\frac{|\bar{\theta}|M}{2}\right)^{3-\frac{n}{2}}
K_{\frac{n}{2}-3} (|\bar{\theta}|M) + 8b_{4}
\left(\frac{|\bar{\theta}|M}{2}\right)^{4-\frac{n}{2}}
K_{\frac{n}{2}-4} (|\bar{\theta}|M)\right]
\end{eqnarray}
This can be seen to coincide completely with the corresponding real part of the
perturbative calculation in \cite{Brandt:2001ud}.

\section{Closed form expressions for real and imaginary parts of the
self-energy}   

Our discussion so far has involved imaginary and real parts of the
self-energy represented in a parametric form (namely, with integrals
over the Feynman parameter). In this section, we will show that the
integration over the Feynman parameter can, in fact, be done exactly
and that both the imaginary as well as the real parts of the
self-energy have closed form expressions in terms of well known
functions. 

To begin with, let us consider the self-energy for the scalar theory
discussed in section {\bf II}. As we have seen there 
(see (\ref{imaginarypi})), the imaginary part of the self-energy is given by
\begin{equation}
{\rm Im}\,\Pi^{\rm (non\hbox{-}planar)} =
\frac{\theta(p^{2})\pi^{\frac{n}{2}+1}}{(2\pi)^{n}} \int_{0}^{1}
dx\,\frac{1}{|M|^{4-n}}\left(\frac{|\bar{\theta}||M|}{2}\right)^
{2-\frac{n}{2}} J_{\frac{n}{2}-2} (|\bar{\theta}||M|)
\end{equation}
where
\begin{equation}
|M| = |(-x(1-x)p^{2})|^{\frac{1}{2}}
\end{equation}
The $x$ integration can, in fact, be done and leads to
\begin{equation}
{\rm Im}\,\Pi^{\rm (non\hbox{-}planar)} =
\frac{\theta(p^{2})\pi^{\frac{n+3}{2}}}{(2\pi)^{n}}\,(p^{2})^{\frac{n-4}{2}}
\left(|\bar{\theta}|(p^{2})^{\frac{1}{2}}\right)^{\frac{3-n}{2}}
J_{\frac{n-3}{2}}
(\frac{|\bar{\theta}|(p^{2})^{\frac{1}{2}}}{2})\label{closedimaginary} 
\end{equation}

The closed form evaluation of the real part of the self-energy, on the
other hand, is slightly more involved. As we have seen in
(\ref{realpi'}),  the real part
of the self-energy can be written as
\begin{equation}
{\rm Re}\,\Pi^{\rm (non\hbox{-}planar)} =
\frac{2\pi^{\frac{n}{2}}}{(2\pi)^{n}}\left(\frac{|\bar{\theta}|}{2}\right)^
{2-\frac{n}{2}} \int_{0}^{1} dx\,(x(1-x))^{\frac{n}{4}-1}
\int_{0}^{\infty} dz\,\frac{1}{z^{2}-p^{2}}\,z^{\frac{n}{2}-1}
J_{\frac{n}{2}-2} (|\bar{\theta}|\sqrt{x(1-x)}z)
\end{equation}
The $x$ integral can be done and leads to
\begin{equation}
{\rm Re}\,\Pi^{\rm (non\hbox{-}planar)} =
\frac{2\pi^{\frac{n+1}{2}}}{(2\pi)^{n}}\left(|\bar{\theta}|
\right)^{\frac{3-n}{2}}  \int_{0}^{\infty}
dz\,\frac{1}{z^{2}-p^{2}}\,z^{\frac{n-3}{2}} J_{\frac{n-3}{2}}
(\frac{|\bar{\theta}|z}{2})
\end{equation}  
Furthermore, the $z$ integration can also be done and the result can,
in fact,  be written
in terms of generalized hypergeometric functions. However, this is not
very useful. Instead, a more useful form is as follows.
\begin{eqnarray}
{\rm Re}\,\Pi^{\rm (non\hbox{-}planar)} & = &
{\rm Re}\,\frac{\pi^{\frac{n+3}{2}}}{2(2\pi)^{n}}
\left(|\bar{\theta}|\right)^{{4-n}}
\frac{(|\bar\theta|^2\,p^{2})^{\frac{n-5}{4}}}{\cos
\frac{(n-3)\pi}{2}}\left[(-1)^{\frac{n-3}{2}} \left({\bf H}_{\frac{3-n}{2}}
(-\frac{|\bar{\theta}|\sqrt{p^{2}}}{2}) - N_{\frac{3-n}{2}}
(-\frac{|\bar{\theta}|\sqrt{p^{2}}}{2})\right)\right.\nonumber\\
 &  & \qquad\qquad \;\;\;\;\;\;\;\;\;\;\;\;\;\;\;\;\;\;\;\;\;
\;\;\;\;\;\;\;\;\;\;\;\;\;\;\;\;\;\;\;\;\;
\;\;\;\;\;\;\;\;\;\;\;
\left. - \left({\bf H}_{\frac{3-n}{2}}
(\frac{|\bar{\theta}|\sqrt{p^{2}}}{2}) - N_{\frac{3-n}{2}}
(\frac{|\bar{\theta}|\sqrt{p^{2}}}{2})\right)\right]\label{struve}
\end{eqnarray}
Here ${\bf H}$ denotes the Struve function while $N$ is the Neumann
function \cite{gradshteyn}
and the principal value prescription is to be understood when 
$p^2> 0$. We note that, in spite of the apparent singularity (coming
from the cosine in the denominator) for $n=2k$, this expression is, in
fact, well behaved for $k>1$ because of the identity
\begin{equation}
{\bf H}_{\frac{3-2k}{2}} (z) = N_{\frac{3-2k}{2}} (z)
\end{equation}
However, for $n=4,6$, which are the cases we are interested in (as we
will see, 
this is what will be needed to discuss the self-energy for QED), it is
nontrivial to extract the finite part from (\ref{struve}). Therefore, in what
follows, we give an alternative closed form expression for the real part
of the self-energy when $n=4,6$.

Let us define
\begin{eqnarray}
S_{1} & = & {\rm Re}\,\Pi^{\rm (non\hbox{-}planar)}_{n=4} =
\frac{2\pi^{\frac{5}{2}}}{(2\pi)^{4}{|\bar{\theta}|}^{{1 \over 2}}} 
\int_{0}^{\infty} dz\,\frac{1}{z^{2}-p^{2}}\,z^{\frac{1}{2}}
J_{\frac{1}{2}} (\frac{|\bar{\theta}|z}{2})\nonumber\\
S_{2} & = & {\rm Re}\,\Pi^{\rm (non\hbox{-}planar)}_{n=6} =
\frac{2\pi^{\frac{7}{2}}}{(2\pi)^{6}|\bar{\theta}|^{\frac{3}{2}}}
\int_{0}^{\infty} dz\,\frac{1}{z^{2}-p^{2}}\,z^{\frac{3}{2}}
J_{\frac{3}{2}} (\frac{|\bar{\theta}|z}{2})
\end{eqnarray}
With the explicit representations for the Bessel functions,
\begin{equation}
J_{\frac{1}{2}} (z) = \sqrt{\frac{2}{\pi z}}\,\sin z,\qquad
J_{\frac{3}{2}} (z) = \sqrt{\frac{2}{\pi z}}\,\left(\frac{\sin z}{z} -
\cos z\right)
\end{equation}
the $z$ integration can be explicitly done and leads to
\begin{eqnarray}
S_{1} & = &
\frac{1}{4\pi^{2}|\bar{\theta}|(p^{2})^{\frac{1}{2}}}\left[\cos
\frac{|\bar{\theta}|(p^{2})^{\frac{1}{2}}}{2}\,(si
\frac{|\bar{\theta}|(p^{2})^{\frac{1}{2}}}{2} + \frac{\pi}{2}) - \sin
\frac{|\bar{\theta}|(p^{2})^{\frac{1}{2}}}{2} ci
\frac{|\bar{\theta}|(p^{2})^{\frac{1}{2}}}{2}\right]\nonumber\\
S_{2} & = &
\frac{1}{16\pi^{3}|\bar{\theta}|^{2}}\left[\frac{2}{|\bar{\theta}|(p^{2})^
{\frac{1}{2}}}\left(\cos
\frac{|\bar{\theta}|(p^{2})^{\frac{1}{2}}}{2}\, (si
\frac{|\bar{\theta}|(p^{2})^{\frac{1}{2}}}{2} +
\frac{\pi}{2}) - \sin \frac{|\bar{\theta}|(p^{2})^{\frac{1}{2}}}{2}\,
ci \frac{|\bar{\theta}|(p^{2})^{\frac{1}{2}}}{2}\right)\right.\nonumber\\
 &  & \qquad
\;\;\;\;\;\;\;\;\;\;\;\;\;\;\;\;\;\;\;
\left. + \left(\cos
\frac{|\bar{\theta}|(p^{2})^{\frac{1}{2}}}{2} ci
\frac{|\bar{\theta}|(p^{2})^{\frac{1}{2}}}{2} + \sin
\frac{|\bar{\theta}|(p^{2})^{\frac{1}{2}}}{2} (si
\frac{|\bar{\theta}|(p^{2})^{\frac{1}{2}}}{2} +
\frac{\pi}{2})\right)\right]\label{S}
\end{eqnarray}
Here, $ci (z), si(z)$ are the well known cosine and sine integrals
defined as \cite{gradshteyn}
\begin{eqnarray}
ci (z) & = & - \int_{z}^{\infty} dt\,\frac{\cos t}{t} = {\bf C} + \ln
z + \int_{0}^{z} dt\,\frac{\cos t -1}{t}\nonumber\\
si (z) & = & - \int_{z}^{\infty} dt\,\frac{\sin t}{t} = -
\frac{\pi}{2} + \int_{0}^{z} dt\,\frac{\sin t}{t}
\end{eqnarray}
where ${\bf C}$ represents Euler's constant. From the series
representation  of the $ci
(z), si(z)$ functions, we obtain in a simple manner the behavior of
the real  parts of the
self-energy as $|\bar{\theta}|\rightarrow 0$, namely,
\begin{eqnarray}
\lim_{|\bar{\theta}|\rightarrow 0}\, S_{1} &\rightarrow & -
\frac{1}{8\pi^{2}}\, \left(\ln
\frac{|\bar{\theta}||p^{2}|^{\frac{1}{2}}}{2} + {\bf C}-1\right)\nonumber\\
\lim_{|\bar{\theta}|\rightarrow 0}\, S_{2} &\rightarrow &
\frac{1}{16\pi^{3}|\bar{\theta}|^{2}}\,\left(1+\frac{|\bar{\theta}|^{2}p^{2}}
{9} -\frac{|\bar{\theta}|^{2}p^{2}}{12}({\bf C} +\ln
\frac{|\bar{\theta}||p^{2}|^{\frac{1}{2}}}{2})\right)\label{limit}
\end{eqnarray}

With these results, we can now discuss the closed form expressions for
the real and the imaginary parts of the self-energy for QED. For
simplicity, we give the results in four dimensions in the Feynman
gauge where  the expressions can be written in terms of the functions
$S_{1},S_{2}$ in (\ref{S}) corresponding to the scalar
self-energy in $n=4,6$ dimensions. After carrying out the $x$
as well as the $z$ integrations, the real and the imaginary parts of
the functions $\Pi_{1},\Pi_{2}$ take the forms (in the Feynman gauge)
\begin{eqnarray}
{\rm Im}\,\Pi_{1}^{\rm (non\hbox{-}planar)} & = &
\frac{4e^{2}\pi^{3}\theta
(p^{2})}{(2\pi)^{4}}\,\sqrt{\frac{\pi}{|\bar{\theta}|(p^{2})^{\frac{1}{2}}}}
\left(J_{\frac{1}{2}} (\frac{|\bar{\theta}|(p^{2})^{\frac{1}{2}}}{2})
- \frac{1}{|\bar{\theta}|(p^{2})^{\frac{1}{2}}}\,J_{\frac{3}{2}}
(\frac{|\bar{\theta}|(p^{2})^{\frac{1}{2}}}{2})\right)\nonumber\\
{\rm Re}\,\Pi_{1}^{\rm (non\hbox{-}planar)} & = & e^{2}\left[
\frac{1}{\pi^{2}|\bar{\theta}|^{2}p^{2}} + 4S_{1} -
\frac{16\pi}{p^{2}} S_{2}\right]\nonumber\\
{\rm Im}\,\Pi_{2}^{\rm (non\hbox{-}planar)} & = &
\frac{2e^{2}\pi^{3}\theta(p^{2})
p^{2}}{(2\pi)^{4}}\,\sqrt{\frac{\pi}{|\bar{\theta}|(p^{2})^{\frac{1}{2}}}}
\left(- J_{\frac{1}{2}}
(\frac{|\bar{\theta}|(p^{2})^{\frac{1}{2}}}{2}) +
\frac{6}{|\bar{\theta}|(p^{2})^{\frac{1}{2}}}\,J_{\frac{3}{2}}
(\frac{|\bar{\theta}|(p^{2})^{\frac{1}{2}}}{2})\right)\nonumber\\
{\rm Re}\,\Pi_{2}^{\rm (non\hbox{-}planar)} & = & e^{2}\left[-
\frac{1}{\pi^{2}|\bar{\theta}|^{2}} - 2p^{2} S_{1} + 48
\pi S_{2}\right]\label{result}
\end{eqnarray}
From the limiting behaviors in (\ref{limit}), it is easy to see that
the imaginary parts of $\Pi_{1},\Pi_{2}$ in (\ref{result}) are well
behaved  when
$|\bar{\theta}|\rightarrow 0$ while the real parts diverge.

\section{Discussion}

In this paper, we have systematically studied and shown that the
dispersion relations hold true for the self-energy in the
non-commutative  $\phi^{3}$
theory as well as in QED in any dimensions. This also explains, as a
manifestation of the IR/UV mixing, how a well behaved imaginary part
of the self-energy develops a divergence structure in the real part as
$|\bar{\theta}|\rightarrow 0$. 

Another interesting aspect of non-commutative
QED is the on-shell behavior of the self-energy. 
Since $\Pi_2$ has no planar contribution, we see from Eq. (\ref{result}) that
the complete structure satisfies
\begin{equation}
\lim_{p^{2}\rightarrow 0}\,{\rm Im}\,\Pi_{2}^{\rm (non\hbox{-}planar)}= 
\lim_{p^{2}\rightarrow 0}\,{\rm Im}\,\Pi_{2} = 0 .
\end{equation}
Similarly, we note from Eq. (\ref{result}) that, for $p^2>0$,
\begin{equation}
\lim_{p^{2}\rightarrow 0}\,{\rm Im}\,\Pi_{1}^{\rm (non\hbox{-}planar)}
= \frac{2e^{2}\pi^{3}}{(2\pi)^{4}}\,\frac{10}{6}
\end{equation}
This does not vanish independently, but for $\Pi_{1}$, there is a
nontrivial planar term \cite{Brandt:2001ud} which gives, for $p^{2}>0$,
\begin{equation}
{\rm Im}\,\Pi_{1}^{\rm (planar)} = -
\frac{2e^{2}\pi^{3}}{(2\pi)^{4}}\, \frac{10}{6}
\end{equation}
so that once again, we see
\begin{equation}
\lim_{p^{2}\rightarrow 0}\,{\rm Im}\,\Pi_{1} = 
\lim_{p^{2}\rightarrow 0}\,{\rm Im}\,\Pi_{1}^{\rm (planar)} +
\lim_{p^{2}\rightarrow 0}\,{\rm Im}\,\Pi_{1}^{\rm (non\hbox{-}planar)} = 0
\end{equation}
These results can be understood simply as follows. As we see
from the explicit forms of the imaginary parts in (\ref{result}), the
combination
$|\bar{\theta}|(p^{2})^{\frac{1}{2}}$ comes together in the arguments
so that $p^{2}\rightarrow 0$ can also be thought of as
$|\bar{\theta}|\rightarrow 0$. However, since the integrand
of the self-energy involves the
factor $(1-\cos\bar{\theta}\cdot k)$, its imaginary part which behaves
smoothly as $|\bar{\theta}|\rightarrow 0$ will vanish in this limit.

The real parts of the self-energy, on the other hand, are
non-vanishing in this limit. In fact, we see from Eqs. (\ref{limit})
and (\ref{result}) that
\begin{equation}
\lim_{p^{2}\rightarrow 0}\,{\rm Re}\,\Pi_{2} =
\lim_{p^{2}\rightarrow 0}\,{\rm Re}\,\Pi_{2}^{\rm (non\hbox{-}planar)}
= \frac{e^{2}}{(4\pi)^{2}}\frac{32}{|\bar{\theta}|^{2}}
\end{equation}
which is a gauge independent result. Furthermore, in the Feynman
gauge,  we have
\begin{equation}
\lim_{p^{2}\rightarrow 0}\,{\rm Re}\,\Pi_{1}^{\rm (non\hbox{-}planar)} =
\frac{2e^{2}}{(4\pi)^{2}}\left[-\frac{10}{6}\left(2{\bf C} + \ln
\frac{|\bar{\theta}|^{2}|p^{2}|}{4}\right)+\frac{28}{9}\right]
\end{equation}
The planar part, on the other hand, behaves like ordinary Yang-Mills
theory and leads to (in the Feynman gauge)
\begin{equation}
\lim_{p^{2}\rightarrow 0}\,{\rm Re}\,\Pi_{1}^{\rm (planar)} = -
\frac{2e^{2}}{(4\pi)^{2}} \left[- \frac{10}{6}\left({\bf C} + \ln
\frac{|p^{2}|}{4\pi\mu^{2}}\right) + \frac{31}{9}\right]
\end{equation}
where $\mu$ represents the renormalization scale. As a result, we see
that the $\ln |p^{2}|$ terms cancel in the complete real part and we
obtain 
\begin{equation}
\lim_{p^{2}\rightarrow 0}\,{\rm Re}\,\Pi_{1} = -
\frac{2e^{2}}{(4\pi)^{2}}\left[\frac{10}{6}\left({\bf C}+ \ln
(\pi|\bar{\theta}|^{2}\mu^{2})\right) + \frac{1}{3}\right]
\end{equation}
There are several things to note from this. First, unlike in ordinary
Yang-Mills theory, the self-energy in non-commutative QED does not exhibit any
logarithmic singularity as $p^{2}\rightarrow 0$. Second, even though the
imaginary parts of $\Pi_{1},\Pi_{2}$
vanish because of the $(1-\cos \bar{\theta}\cdot k)$ factors in the
integrand, the real parts do not, which can be understood 
as a consequence of the large energy behavior of the dispersion integral.

Let us discuss this feature in some more detail. First, we note from
Eq. (\ref{result}) that, for a fixed $p^{2}$,
\begin{equation}
\lim_{|\bar{\theta}|\rightarrow 0}\,{\rm Im}\,\Pi_{2} =
\lim_{|\bar{\theta}|\rightarrow 0}\,{\rm Im}\,\Pi_{2}^{\rm
(non\hbox{-}planar)} \sim |\bar{\theta}|^{2} (p^{2})^{2} \rightarrow 0
\end{equation}
The dispersion relation, on the other hand, leads to
\begin{eqnarray}
\lim_{|\bar{\theta}|\rightarrow 0}\,{\rm Re}\,\Pi_{2} &
= & \lim_{|\bar{\theta}|\rightarrow 0}\, \frac{2}{\pi} \int_{0}^{\infty}
d\omega\,\frac{\omega\,\theta(\omega^{2}-\vec{p}^{2})}{\omega^{2}-(p^{0})^{2}}
\,{\rm Im}\,\Pi_{2} (\omega,\vec{p},|\bar\theta|)\nonumber\\
 &\sim & \int_{0}^{\infty}
d\omega\,\frac{\omega\,\theta(\omega^{2}-\vec{p}^{2})}{\omega^{2}-(p^{0})^{2}}
\,|\bar{\theta}|^{2} (\omega^{2}-\vec{p}^{2})^{2}
\end{eqnarray}
Thus, we see that, although for any finite $\omega$, the integrand
vanishes as $|\bar{\theta}|\rightarrow 0$, it is, in fact, strongly
divergent as $\omega\rightarrow \infty$ leading to a nontrivial real
part that diverges as $\frac{1}{|\bar{\theta}|^{2}}$ for small
$|\bar{\theta}|$. 

We can analyze the corresponding behavior for $\Pi_{1}$ as well. Here,
although a new feature develops since
$\Pi_{1}^{\rm (planar)}$ is divergent and satisfies a subtracted
dispersion relation, the analysis is completely parallel and leads to the result
that the real part of $\Pi_{1}$ diverges logarithmically as
$|\bar{\theta}|\rightarrow 0$. A similar analysis can also be carried
out for non-commutative \un theory.
\vskip 1cm

\noindent{\bf Acknowledgment:}

We would like to thank Prof. J. C. Taylor for helpful comments.
This work was supported in part by US DOE Grant number DE-FG
02-91ER40685, by CNPq and FAPESP, Brasil.

\appendix
\section{Calculation of imaginary parts}

In this section, we will discuss briefly how the imaginary parts of
the self-energy are calculated. We note that, for $p^{0}>0$, only one
of the delta functions contributes and, in this case, we have
\begin{eqnarray}
{\rm Im}\,\Pi^{\rm (non\hbox{-}planar)} & = & \frac{\pi}{4
(2\pi)^{n-1}} \int d^{n-1}k\,\frac{\cos\bar{\theta}\cdot
k}{|\vec{k}||\vec{k}+\vec{p}|} \,\delta
(p^{0}-|\vec{k}|-|\vec{k}+\vec{p}|)\nonumber\\
 & = & \frac{\pi}{4 (2\pi)^{n-1}} \int
k^{n-2}dk\,(\sin\theta_{1})^{n-3}d\theta_{1}\,(\sin\theta_{2})^{n-4}
d\theta_{2}\cdots d\theta_{n-2}\,\frac{\cos\bar{\theta}\cdot k}{k
|\vec{k}+\vec{p}|}\,\delta (p^{0}-k-|\vec{k}+\vec{p}|)
\end{eqnarray}
Here, we have identified $k=|\vec{k}|$ as the radial component of the momentum
vector. If we next identify $\vec{p}$ to lie along the $(n-1)$-th axis
and $\vec{\bar{\theta}}$ to lie along the $(n-2)$-th axis, we have
\begin{equation}
\vec{k}\cdot\vec{p} = k|\vec{p}|\cos\theta_{1},\qquad
\bar{\theta}\cdot k = |\bar{\theta}|k\sin\theta_{1}\cos\theta_{2}
\end{equation}

The delta function can now be used to do the $\theta_{1}$
integration. In fact, the delta function determines
\begin{equation}
\cos\theta_{1} = \frac{1}{2k|\vec{p}|}\,(p^{2} - 2kp^{0})
\end{equation}
and limits the range of the $k$ integration to
$k_{min}=\frac{p^{0}-|\vec{p}|}{2}\leq k \leq
\frac{p^{0}+|\vec{p}|}{2}=k_{max}$. This also shows that the imaginary
part is nontrivial only for $p^{2}\geq 0$. Doing the $\theta_{1}$
integral and shifting $k\rightarrow k-k_{min}$, we obtain,
\begin{eqnarray}
{\rm Im}\,\Pi^{\rm (non\hbox{-}planar)} & = & \frac{\theta(p^2)\pi}{4
(2\pi)^{n-1}} \int (\sin\theta_{3})^{n-5}d\theta_{3}\cdots
d\theta_{n-2}\nonumber\\
 &  & \times \int_{0}^{k_{max}-k_{min}}
\frac{dk}{|\vec{p}|}\,(\sin\theta_{2})^{n-4}d\theta_{2}\,\left(\frac{(p^{2}
(k_{max}-k_{min}-k)k)^{\frac{1}{2}}}{|\vec{p}|}\right)^{n-4}\nonumber\\
 &  & \times  \cos
\left(\frac{|\bar{\theta}|(p^{2}(k_{max}-k_{min}-k)k)^{\frac{1}{2}}
\cos\theta_{2}}{|\vec{p}|}\right)
\end{eqnarray}
Rescaling
\begin{equation}
k = x (k_{max}-k_{min}) = x |\vec{p}|
\end{equation}
the imaginary part becomes
\begin{eqnarray}
{\rm Im}\,\Pi^{\rm (non\hbox{-}planar)} & = & \frac{\theta(p^2)\pi}{4
(2\pi)^{n-1}} \int (\sin\theta_{3})^{n-5}d\theta_{3}\cdots
d\theta_{n-2}\nonumber\\
 &  & \times \int_{0}^{1} dx\,(x(1-x)p^{2})^{n-4}
(\sin\theta_{2})^{n-4}d\theta_{2}\,\cos
(|\bar{\theta}|(x(1-x)p^{2})^{\frac{1}{2}}\cos\theta_{2})\nonumber\\
 & = & \frac{\pi^{\frac{n}{2}+1}}{(2\pi)^{n}} \int_{0}^{1}
dx\,\frac{1}{|M|^{4-n}}
\left(\frac{|\bar{\theta}||M|}{2}\right)^{2-\frac{n}{2}}
J_{\frac{n}{2}-2} (|\bar{\theta}||M|)
\end{eqnarray}
where $|M|$ is defined earlier in Eq. (\ref{M}). We take this
opportunity to correct a typographical error in Eq. (61) of 
ref. \cite{Brandt:2001ud}. The imaginary part of the modified Bessel
function satisfies, for $p^2>0$, the relation
\be
{\rm Im}\,{K_l(|\bar\theta|\, M)\over (|\bar\theta|\, M)^l} =
(-1)^l {\pi\over 2} \,{J_l(|\bar\theta|\, |M|)\over (|\bar\theta|\, |M|)^l},
\ee
which can also be used to calculate the imaginary part of the
self-energy from the result in \cite{Brandt:2001ud}.


\end{document}